\documentclass[12pt]{article}

\input epsf
\def\beq{\begin{eqnarray}}
\def\eeq{\end{eqnarray}}

\def\dd{{\delta}}

\def\ep{\epsilon}
\def\mpl{M_{\rm Pl}}
\def\t{\tilde}
\def\tg{\tilde g}

\def\hg{\hat g}
\def\hh{\hat h}

\def\lsim{\mathrel{\rlap{\lower3pt\hbox{\hskip0pt$\sim$}}
     \raise1pt\hbox{$<$}}}         %less than or approx. symbol
\def\gsim{\mathrel{\rlap{\lower4pt\hbox{\hskip1pt$\sim$}}
     \raise1pt\hbox{$>$}}}         %greater than or approx. symbol

\newcommand{\superscript}[1]{\ensuremath{^{\textrm{#1}}}}

\newcommand{\st}[0]{\superscript{st}}
\newcommand{\nd}[0]{\superscript{nd}}
\newcommand{\rd}[0]{\superscript{rd}}

\newcommand{\comment}[1]{}
\usepackage{amsmath}
\usepackage{amsfonts}
\usepackage{verbatim}
\usepackage{graphicx}
\usepackage{subfigure}
\usepackage{amssymb}
\usepackage[T1]{fontenc}
\usepackage{calligra}

\renewcommand{\comment}[1]{}

\begin{document}

\begin{titlepage}

\thispagestyle{empty}

\begin{flushright}
{NYU-TH-11/04/59}
\end{flushright}
\vskip 0.9cm

\centerline{\Large \bf Generalized Framework for Auxiliary Extra Dimensions}
\vskip 0.2cm
\centerline{\Large \bf }                    

\vskip 0.7cm
\centerline{\large Lasha Berezhiani and Mehrdad Mirbabayi}
\vskip 0.3cm
\centerline{\em Center for Cosmology and Particle Physics, Department of Physics,}
\centerline{\em New York University, New York, 
NY  10003, USA}

\vskip 1.9cm

\begin{abstract}

The theory of gravity with an auxiliary extra dimension is known to give the ghost-free cubic completion of the Fierz-Pauli mass term in the decoupling limit. Our work generalizes the boundary condition in the auxiliary dimension that avoids ghosts order-by-order,  and to all orders, in the decoupling limit. Furthermore, we extend the formalism  to the case of many auxiliary dimensions, and
we  show that the multi-dimensional extension  with the rotationally invariant boundaries of the bulk, is  equivalent to the model with a single auxiliary dimension. The above  constructions require the appropriate adjustment of the boundary condition, which we discuss in detail. The other possible extension of the original model by the Gauss-Bonnet term is studied as well.

\end{abstract}

\vspace{3cm}

\end{titlepage}

\newpage

\section{Introduction}

The theory of massive gravity which once had been thought to be ruled out because of propagation of ghost 
at non-linear level \cite{BD,Creminelli}, was recently revitalized in \cite{Giga_FP,Giga_Resum}, where it 
was observed that there is a two parameter family of generalizations of Fierz-Pauli mass term \cite{FP}, 
which is free from ghostlike instability at least in the decoupling limit, and also in the full theory
at least in the quartic order in  nonlinearities \cite {Giga_Resum}.  For further interesting studies
of this model see \cite {Koyama1+2,Nieuwenhuizen,HassanRosen,dato}.
This on the other hand strongly motivates a quest for underlying setups that 
would naturally explain or at least elegantly reproduce the unusual 
structures emerged in \cite{Giga_FP,Giga_Resum}.

The theories of gravity supplemented with an auxiliary extra dimension (AED) \cite{GG,Claudia}, 
inspired by the DGP brane-world gravity model \cite {DGP}, appeared as a promising step in this 
direction as they correctly predicted a cubic ghost-free completion of Fierz-Pauli \cite{Giga_tuned}. 
In these models while all matter fields live on a 4 dimensional brane, the metric has been extended to 
an extra dimension $-1<u<1$, and is called $\tg_{\mu\nu}(x,u)$. The brane is located at $u=0$ and a 
$\mathbf Z_2$ symmetry is imposed on the fields, $\tg_{\mu\nu}(x,u)=\tg_{\mu\nu}(x,-u)$, using which 
the graviton's Lagrangian looks like
\beq
\label{lagr}
{\cal L}=\mpl^2\sqrt{-g}R-\mpl^2m^2\int_0^1du\sqrt{-\tg}(k_{\mu\nu}^2-k^2)\,,
\eeq
where the first term is the Einstein-Hilbert Lagrangian on the brane as a function of $g_{\mu\nu}(x)=\tg_{\mu\nu}(x,u=0)$, while $k_{\mu\nu}=\frac{1}{2}\partial_u\tg_{\mu\nu}$, $k=\tg^{\mu\nu}k_{\mu\nu}$, and all indices in the second term are contracted using inverse extended metric $\tg^{\mu\nu}$. The coordinate $u$ is called an auxiliary dimension because after choosing a second boundary condition, say at $u=1$, $\tg_{\mu\nu}(x,u)$ is algebraically determined in terms of $g_{\mu\nu}$ and the second term in \eqref{lagr} describes just a potential for the induced metric on the brane. 

Of course the choice of this second boundary condition is by no means unique at the level of auxiliary extra dimension: while $\tg_{\mu\nu}(x,u=1)=\eta_{\mu\nu}$ was originally chosen to describe a particular completion of Fierz-Pauli massive gravity, nothing prevents us from considering a more general boundary condition to accommodate a larger family of completions. In this way we have written the most generic graviton's potential in terms of a geometrical construct that naturally arises in higher dimensional theories of gravity, where $k_{\mu\nu}$ is taken to be the extrinsic curvature.

The abovementioned generalization of the boundary condition is straightforward and will be the subject of \S2, where we show that one can consistently take $\tg_{\mu\nu}(x,u=1)$ a generic function of $g_{\mu\nu}$, the 4D-brane metric. In \S3 we perform the dimensional reduction to obtain the 4D potential for graviton, and explicitly show that in the original choice of boundary condition the absence of the ghost does not persist at higher than cubic order (in agreement with \cite{Rachel}). Thereafter a recipe will be provided to reproduce the potential of \cite{Giga_FP} by a proper choice of the boundary condition, ensuring the stability of the full theory in the decoupling limit. In \S4 we consider the generalization of the model to higher extra dimensions and prove that in the most tractable case with a spherically symmetric boundary condition in the bulk, one and many extra dimensions are equivalent. 

We further provide a simple explanation for the absence of the cubic ghost and the presence of higher order ones in the original choice of boundary condition ($\tg_{\mu\nu}(x,u=1)=\eta_{\mu\nu}$), in appendix A. In appendix B we address the possibility of modifying the bulk action by inclusion of other geometrical constructs so as to avoid the adjustment of the boundary condition. It is shown there that the most natural candidate which is (the auxiliary version of) Gauss-Bonnet term, as the only ``Lovelock'' term other than Einstein-Hilbert in 5D for which the variation principle is well defined, cannot do much better than the original action in predicting right coefficients. Nevertheless one can build a ghost-free theory by adding higher powers of $k_{\mu\nu}$ with specially tuned coefficients. Finally, in appendix C we give the most general boundary condition which gives healthy theory at quartic order.

%%%%%%%%%%%%%%%%%%%%%%%%%%%%%%%%%%%%%%%%%%%%%%%%%%%%%%%%%%%%%%

\section{Generalized Boundary Conditions}

We begin this letter by considering the generalization of the boundary conditions for \eqref{lagr}. For convenience we will assume the location of the boundaries to be at $u=0,1$ rather than $u=\pm 1$. As usual we fix  $\tilde{g}_{\mu\nu}(x,u=0)=g_{\mu\nu}(x)$, while the choice at $u=1$ is in principle arbitrary since this boundary does not have an intrinsic dynamics and the values of the fields on it are completely determined by imposed conditions. Hence, the most general boundary conditions would be 
\beq
\tilde{g}_{\mu\nu}(x,u=0)=g_{\mu\nu}(x) \quad \text{and} \quad \tilde{g}_{\mu\nu}(x,u=1)=\hg_{\mu\nu}(g_{\alpha\beta}),
\label{bound}
\eeq
where $\hat g_{\mu\nu}-\eta_{\mu\nu}$ has to vanish when $g_{\mu\nu}=\eta_{\mu\nu}$, if the theory is to describe massive gravity around Minkowski. The variation of the action \eqref{lagr} then generalizes to 
\beq
\dd S=M^2_{pl}\int d^4x\left\{ \sqrt{-g} G_{\mu\nu} \dd g^{\mu\nu}\vert_{u=0}-m^2\left.\left[ \sqrt{-\tilde{g}}(k^{\mu\nu}-\tg^{\mu\nu}k)\dd \tg_{\mu\nu} \right]\right\vert_{u=0} ^{u=1} \right.\nonumber \\
\left. -m^2\int_0^1 du \sqrt{-\tilde{g}} \left[ \partial_u k_{\mu \nu}-\tg _{\mu\nu}\partial_u k-\frac{1}{2}\tg_{\mu\nu}(k_{\alpha\beta}^2+k^2)-2k_{\mu}^{\alpha}k_{\alpha\nu}+kk_{\mu\nu} \right]\dd \tg ^{\mu\nu} \right\},
\eeq
from which it follows that the bulk ($u>0$) equation of motion is 
\beq
 \partial_u k_{\mu \nu}-\tg _{\mu\nu}\partial_u k=\frac{1}{2}\tg_{\mu\nu}(k_{\alpha\beta}^2+k^2)+2k_{\mu}^{\alpha}k_{\alpha\nu}-kk_{\mu\nu},
 \label{bulk}
\eeq
the same as in the case with field independent boundary condition at $u=1$. The brane equation of motion on the other hand gets slightly modified to
\beq
\sqrt{-g}G_{\mu\nu}-m^2 \left\{ \left[ \sqrt{-\tg} (k_{\mu\nu}-\tg_{\mu\nu} k) \right]_0-\left[ \sqrt{-\tg} (k^{\alpha\beta}-\tg^{\alpha\beta} k) \right]_1\frac{\partial \hg_{\alpha\beta}}{\partial g_{\rho\sigma}}\hg_{\mu\rho}\hg_{\nu\sigma}\right\}=0.\nonumber
\eeq
Notice that this differs from the 4D equation of \cite{GG} by the last term, which arises  because of the non-vanishing $\dd \tg _{\mu\nu}\vert_{u=1}$\footnote{It should be mentioned that this does not make the variation principle ill-defined.}.

An equivalent and often simpler way of analyzing the AED models is to integrate over the $u$-coordinate using the solution to the bulk equation \eqref{bulk} with the boundary condition \eqref{bound} in order to obtain an effective 4-dimensional potential for graviton. This was the strategy pursued in \cite{Giga_tuned} for the special case of $\tg_{\mu\nu}(x,u=1)=\eta_{\mu\nu}$, treating $h_{\mu\nu}$ perturbatively, but the generalization is straightforward. One simply needs to write
\beq
g_{\mu\nu}=\eta_{\mu\nu}+h_{\mu\nu}=\hat g_{\mu\nu}+\hat h_{\mu\nu}\,, \quad \text{with} \quad \hat{h}_{\mu\nu}\equiv \eta_{\mu\nu}+h_{\mu\nu}-\hg_{\mu\nu},
\eeq
and expand all $\tg_{\mu\nu}$ around $\hat g_{\mu\nu}$ so that all computations go through identically if the following replacements are made:
\beq
\label{replace}
\eta_{\mu\nu}\to\hat g_{\mu\nu}\,,\qquad\text{and}\qquad h_{\mu\nu}\to\hat h_{\mu\nu}\,.
\eeq
Therefore one just needs to take the original 4D effective potential and make the substitution \eqref{replace} (including the overall factor $\sqrt{-\det(\eta)}=1\to\sqrt{-\det(\hat g)}$).

%%%%%%%%%%%%%%%%%%%%%%%%%%%%%%%%%%%%%%%%%%%%%%%%%%%%%%%%%%%%%%%%%%%

\section{\label{quartic} Quartic Ghost and a Remedy}

Perhaps the simplest method to check models of massive gravity against Boulware-Deser ghost \cite{BD} is to take the routes of refs. \cite{Giga_tuned} and \cite{Lasha}, namely, to write the graviton potential in the unitary gauge, expand it around the Minkowski background and compare it with the unique set of conceivably ghost-free potentials constructed in \cite{Giga_FP}. Obviously to perform this comparison the latter should also be re-expanded in the unitary gauge where no St\"uckelberg fields are present.

Following the logic of the previous section, in order to perform the dimensional reduction perturbatively in $\hat{h}_{\mu\nu}$, we define
\beq
\label{h_tilde}
\t h_{\mu\nu}(x,u)=\tg_{\mu\nu}(x,u)-\hg_{\mu\nu}=H^{(1)}_{\mu\nu}(x,u)+H^{(2)}_{\mu\nu}(x,u)+\dots\,,
\eeq
where $H^{(n)}_{\mu\nu}$ is of n$^{\text{th}}$ order in $\hh_{\mu\nu}$ and except $H^{(1)}_{\mu\nu}$ which is given by
\beq
\label{H1}
H^{(1)}_{\mu\nu}(x,u)=(1-u)\hh_{\mu\nu}(x)\,,
\eeq
all the rest vanish both at $u=0$ and $u=1$, so that $\tg_{\mu\nu}(x,u)$ satisfies the desired boundary conditions at each order. This property of higher order bulk solutions together with the absence of second or higher derivatives of fields in the bulk action ensures that to obtain the n$^{\text{th}}$ order graviton potential one only needs to solve $\t h_{\mu\nu}$ up to (n-2)\nd order. To wit note that the action \eqref{lagr} starts from quadratic in $\t h_{\mu\nu}$ which thus contains the n$^{\text{th}}$ order term of the schematic form
\beq
\int_0^1 du\;\partial_uH_{\mu\nu}^{(n-1)}\partial_uH_{\mu\nu}^{(1)}= \left. H_{\mu\nu}^{(n-1)}\partial_uH_{\mu\nu}^{(1)}\right \vert_0^1-\int_0^1 du \;H_{\mu\nu}^{(n-1)}\partial_u^2 H_{\mu\nu}^{(1)}\,,
\eeq
but the first term on the r.h.s. vanishes because of the vanishing of $H^{(n-1)}_{\mu\nu}(u)$ on the boundaries, and the second term because of the linear equation of motion which is satisfied by $H^{(1)}_{\mu\nu}(u)$.

In the case of field-independent boundary value $\hat g_{\mu\nu}=\eta_{\mu\nu}$, the comparison with the tuned polynomial was carried out in \cite{Giga_tuned} up to cubic order yielding agreement with one class of (decoupling limit) ghost-free potentials characterized by $c_3=1/4$. Here we show that this agreement breaks down at quartic order and in the appendix A using a simple covariantization we explain why it could have been expected. We also find the most general modification of boundary condition that give rise to the ghost free 4D action up to quartic order. This is achieved by perturbative adjustment of the boundary value at $u=1$, hence one can continue this process to all orders.

As already mentioned, to obtain fourth order 4D effective potential one needs only up to the second order bulk solution:
\beq
\label{H2}
H^{(2)}_{\mu\nu}=\frac{1}{2}u(1-u)\left[\frac{1}{12}\eta_{\mu\nu}(\hh_{\sigma\rho}^2-\hh^2)-\hh_\mu^\sigma \hh_{\sigma\nu}+\frac{1}{2}\hh\hh_{\mu\nu}\right]\,,
\eeq
which after some work yields the following unitary-gauge potential
\beq
\label{V_4}
V^{(4)}(h_{\mu\nu})&=&\sqrt{-\hg}\left(\hh_{\mu\nu}^2-\hh^2-\hh_{\mu\nu}^3+\frac{5}{4}\hh\hh_{\mu\nu}^2-\frac{1}{4}\hh^3\right.\nonumber\\
&&\left.+\frac{11}{12}\hh_{\mu\nu}^4-\frac{11}{12}\hh\hh_{\mu\nu}^3-\frac{53}{144}(\hh_{\mu\nu}^2)^2+\frac{29}{72}\hh^2\hh_{\mu\nu}^2-\frac{5}{144}\hh^4\right)\,,
\label{general}
\eeq
with indices contracted by matrix $\hg^{\mu\nu}$, inverse to $\hg_{\mu\nu}$. This expression should be compared to 
\beq
\label{V(H)}
V^{(4)}_{\rm{tuned}}(H_{\mu\nu})&=&\sqrt{-g}\left[H_{\mu\nu}^2-H^2+c_1H_{\mu\nu}^3+c_2HH_{\mu\nu}^2+c_3H^3\right.\nonumber\\
&&\left. +d_1H_{\mu\nu}^4+d_2HH_{\mu\nu}^3+d_3(H_{\mu\nu}^2)^2+d_4H^2H_{\mu\nu}^2+d_5H^4\right]\,,
\eeq
where $H_{\mu\nu}$, which is a covariant version of $h_{\mu\nu}=g_{\mu\nu}-\eta_{\mu\nu}$, reduces to $h_{\mu\nu}$ in the unitary gauge, indices are now contracted with the full metric $g^{\mu\nu}$ and all the coefficients are expressed in terms of $c_3$ and $d_5$ as given in \cite{Giga_FP}. 

Concentrating on the case of constant boundary $\hg_{\mu\nu}=\eta_{\mu\nu}$, where we know that $c_3=1/4$ guaranties up to 3\rd order agreement between \eqref{V_4} and \eqref{V(H)}, and, expanding the latter in the unitary gauge around Minkowski background we get
\beq
V^{(4)}_{\rm{tuned}}(h_{\mu\nu})&=&h_{\mu\nu}^2-h^2-h_{\mu\nu}^3+\frac{5}{4}hh_{\mu\nu}^2-\frac{1}{4}h^3\nonumber\\
&&+d_1h_{\mu\nu}^4+d_2hh_{\mu\nu}^3+d_3(h_{\mu\nu}^2)^2+d_4h^2h_{\mu\nu}^2+d_5h^4\,.
\eeq
However this can never coincide with \eqref{V_4} for any value of $d_5$ and consequently the theory suffers from a ghost at quartic order. This can most easily be seen by introducing St\"uckelberg fields in terms of which there remains a quartic self-interaction of the helicity-0 part $\pi$ of the schematic form
\beq
\label{pi_4}
\mpl^2m^2(\partial\partial\pi)^4 = \frac{1}{\mpl^2m^6}(\partial\partial\pi_c)^4\equiv \frac{1}{\Lambda_4^8}(\partial\partial\pi_c)^4 \,,
\eeq
where $\pi_c=\mpl m^2 \pi\equiv\Lambda_3^3\pi$ is the canonically normalized field. In terms of interactions of $\pi$ the agreement with the adjusted potential to 3\rd order translates into the absence of the cubic interactions:
\beq
\label{pi_3}
\frac{1}{\mpl m^4}(\partial\partial\pi_c)^3\equiv \frac{1}{\Lambda_5^5}(\partial\partial\pi_c)^3 \,,
\eeq
which are generically present in theories of massive gravity \cite{AGS}. Consequently \eqref{pi_4} is the most strongly coupled interaction in the theory or equivalently the scale $\Lambda_4=(\mpl m^3)^{1/4}$ is the smallest mass scale by which any interaction in this model may be suppressed. We can therefore define a decoupling limit 
\beq
\mpl\to\infty,\qquad m\to 0,\qquad \Lambda_4-\rm{fixed},
\eeq
in which \eqref{pi_4} is the only interaction that survives and it propagates ghosts lighter than the cutoff around any reasonable astrophysical background (as shown in \cite{Creminelli} for the similar case of the cubic interaction \eqref{pi_3}). The fact that the action contains only a finite number of terms leaves no ambiguity in the instability of the model since no resummation may be invoked to remove the ghost.

Having discussed the problems of the original model let us comment on \eqref{general} with general boundary condition. A similar analysis shows that ghost can be avoided by order-by-order adjustment of the boundary condition
\beq
\hg_{\mu\nu}&=&\eta_{\mu\nu}+b^{(1)}_1\eta_{\mu\nu}h+b^{(1)}_2h_{\mu\nu}\nonumber \\
&+&b^{(2)}_1\eta_{\mu\nu}h_{\alpha\beta}^2+b^{(2)}_2\eta_{\mu\nu}h^2+b^{(2)}_3h_{\mu\nu}h+b^{(2)}_4[h^2]_{\mu\nu}\nonumber \\
&+&b^{(3)}_1\eta_{\mu\nu}h_{\alpha\beta}^3+b^{(3)}_2\eta_{\mu\nu}hh_{\alpha\beta}^2+b^{(3)}_3\eta_{\mu\nu}h^3+b^{(3)}_4h_{\mu\nu}h_{\alpha\beta}^2\nonumber\\
&+&b^{(3)}_5h_{\mu\nu}h^2+b^{(3)}_6[h^2]_{\mu\nu}h+b^{(3)}_7[h^3]_{\mu\nu}\nonumber \\
&&\ldots
\label{expansion}
\eeq
The requirement of the Fierz-Pauli structure constrains $b^{(1)}_{1,2}$, the absence of the cubic Boulware-Deser ghost relates the quadratic coefficients $b^{(2)}_{1,\ldots,4}$ to each other and so on. The most general healthy coefficients of the above expansion are explicitly given in appendix C. Here, on the other hand, we give just one of the simplest expressions for $\hg_{\mu\nu}$ which cures the instability of the theory in quartic order
\beq
\hg_{\mu\nu}=\eta_{\mu\nu}+\frac{1}{96}\eta_{\mu\nu}h_{\alpha\beta}^3-\frac{7}{432}\eta_{\mu\nu}hh_{\alpha\beta}^2+\frac{5}{864}\eta_{\mu\nu}h^3-\frac{17}{288}h_{\mu\nu}h_{\alpha\beta}^2+\frac{11}{96}[h^3]_{\mu\nu}.
\eeq
The theory with this choice of the boundary condition, once reduced to 4D gives the 4$^{\text{th}}$ order potential of \cite{Giga_FP} corresponding to $c_3=1/4$ and $d_5=0$.

We would like to stress that our approach is perturbative in contrast to \cite{Rachel}, where authors have performed the $u$-integral exactly  and obtained nonlinear 4D action as a function of $\hg_{\mu\nu}$ (in their notation the metric at $u=1$ is labeled as $f_{\mu\nu}$). The advantage of that framework is that one may try to find an exact ghost-free boundary condition by equating $F(g^{\mu\nu}\hg_{\nu\alpha})$ of \cite{Rachel} to the ghost-free potential $\mathcal{U}_{\text{gh-fr}}(g_{\mu\nu},\eta_{\mu\nu})$ of \cite{Giga_Resum}. However, because of the transcendental nature of the equation
\beq
&&\left[\text{det}(g^{\mu\alpha}\hg_{\alpha\nu})\right]^{1/2}-2\left[\text{det}(g^{\mu\alpha}\hg_{\alpha\nu})\right]^{1/4}\times\nonumber\\
&&\text{cosh}\left( \frac{1}{2\sqrt{3}}\sqrt{\text{Tr}[\text{ln}(g^{\mu\alpha}\hg_{\alpha\nu})]^2-\frac{1}{4}[\text{Tr ln}(g^{\mu\alpha}\hg_{\alpha\nu} ) ]^2} \right)+1=\frac{1}{3}\mathcal{U}_{\text{gh-fr}},
\eeq
one seems to be forced to do the perturbative analysis.

%%%%%%%%%%%%%%%%%%%%%%%%%%%%%%%%%%%%%%%%%%%%%%%%%%%%%%%%%
\section{Generalization to Multi-D}

Another natural generalization of AED is to consider a multi-dimensional auxiliary space instead of a single dimensional one. In this section we investigate the special case when there is a rotationally invariant condition on the boundaries of the $d$-dimensional bulk, and show that it gives identical 4D effective action as the original AED model.

Using the cartesian coordinates $y_a$, where $a = 1,2,\dots,d$, the action of the extra dimensions generalizes to 
\beq
\label{S_d_dim}
S_d=\int d^d y \sqrt{-\tilde{g}}(k_{a\mu \nu}^2-k_a^2),
\eeq
where $k_{a\mu\nu}\equiv\frac{1}{2}\partial_a\tg_{\mu\nu}$ and summation on repeated indices is implied. We are willing to impose a spherically symmetric boundary condition, however, one should be cautious in this case since the solutions of the Laplace equation are singular at the origin. This is a generic feature of frameworks with co-dimension >1 branes \cite{derham-tolley}, and it is well known that the singularity can be regularized by assigning a finite width $\epsilon$ to the brane. It suffices to take the radial coordinate, defined as $r\equiv\sqrt{y_ay_a}$, to range in the interval $r\in [\epsilon,1]$. Therefore the region $r<\ep$ is excluded from the integral in \eqref{S_d_dim}, and the boundary conditions are modified to
\beq
\tg_{\mu\nu}(x,r=\ep)=g_{\mu\nu},\qquad\tg_{\mu\nu}(x,r=1)=\hat g_{\mu\nu}\,.
\eeq
From this boundary condition it follows that the solution to the bulk equations of motion will also be spherically symmetric and therefore \eqref{S_d_dim} can be simplified considerably by going to the spherical coordinates. Integrating over angular variables and dropping an unimportant normalization constant which can always be absorbed in the definition of the graviton's mass we obtain
\beq
\label{r_d}
S_d=\int_\ep^1 r^{d-1}dr \sqrt{-\tilde{g}}\left[(k_{r \mu \nu})^2-k_r^2\right],
\eeq
with $k_{r\mu\nu}\equiv\frac{1}{2}\partial_r\tg_{\mu\nu}$. However by a change of the variable of integration to 
\beq
du=\frac{dr}{r^{d-1}}\,,
\eeq
and a rescaling such that $u$ ranges in $[0,1]$, the action \eqref{r_d} transforms to that of single extra dimension model \eqref{lagr}. Hence the spherically symmetric multi-D analog is equivalent to the original model with one extra dimension. However, this equivalence does not generically persist the modifications of the bulk action, see appendix B for more details.

%%%%%%%%%%%%%%%%%%%%%%%%%%%%%%%%%%%%%%%%%%%%%%%%%%%%%%%%%%%%%%%%%%%%%%%

\section{Acknowledgements}
We are grateful to Gregory Gabadadze for his guidance, useful discussions and support. LB and MM are supported respectively by MacCracken and James Arthur Graduate Fellowships at NYU.

\renewcommand{\theequation}{A-\Roman{equation}}
\setcounter{equation}{0} 

\section*{Appendix A. Covariantization by $N_\mu$}
%\section{Covariantization by $N_\mu$}

As first pointed out in \cite{Claudia}, AED models with fixed (e.g. $\eta_{\mu\nu}$) boundary condition at $u=1$ have their own natural candidate for playing the role of St\"uckelberg fields, namely the ADM shift vector $N_\mu$. Consider Einstein-Hilbert action in 5 dimension with metric $G_{MN}$ and decompose it on spatial slices of constant 5$^{\text{th}}$ coordinate using ADM parameters: lapse $N=(G^{55})^{-1/2}$, shift $N_\mu = G_{5\mu}$, and 4D metric $\tg_{\mu\nu}\equiv G_{\mu\nu}$
\beq
S=2M_5^2 \int d^4x du \sqrt{-\det(\tg_{\mu\nu})}N [R^{(4)}(\tg) - K^{\mu\nu}K_{\mu\nu}+ K^2]\,,
\eeq
where
\beq
K_{\mu\nu}=\frac{1}{2N}(\partial_u \tg_{\mu\nu}- D_\mu N_\nu - D_\nu N_\mu)\,,
\eeq
all indices are raised using $(\tg^{-1})^{\mu\nu}\equiv\tg^{\mu\nu}$, and $D_\mu$ is the covariant derivative with respect to the 4D metric. The action is invariant under the full 5D diffeomorphisms and in particular the subclass of $u$-dependent 4D diffeomorphisms
\beq
\label{diff}
u\to u'=u\,,\qquad x^\mu \to {x'}^\mu(x,u)\,,
\eeq
under which $N$ and $R^{(4)}$ behave as scalars which implies that $K_{\mu\nu}$ must transform as a covariant tensor. 

Likewise, we can restore $u$-dependent diffeomorphism by replacing $k_{\mu\nu}$ in the action of AED with $K_{\mu\nu}=(\partial_u \tg_{\mu\nu}/2- D_{(\mu }N_{\nu)})$ and stipulating that $N_\mu$, which is now some auxiliary field, transforms the same way as the 5D gravity's shift vector does, namely
\beq
N^\mu \to N^\nu \frac{\partial x'^\mu}{\partial x^\nu} +\partial_u x'^\mu\,.
\eeq
Having restored this class of diffeomorphisms the action is now invariant if one reparametrizes the 4D metric $g_{\mu\nu}$ on $u=0$ brane but keep the other boundary fixed at $\tg_{\mu\nu}(x,u=1)=\eta_{\mu\nu}$. In other words $N_\mu$ covariantizes the 4D effective Lagrangian, and can be regarded as the St\"uckelberg field. This covariantization fails to work in the more general case where the boundary condition at $u=1$ ($\hat g_{\mu\nu}$) depends on $g_{\mu\nu}$ because now the reparametrization at $u=0$ brane changes the $\hat g_{\mu\nu}$ but not necessarily in the manner of a coordinate transformation.

In terms of $N_\mu$ it is easy to understand the absence of the cubic ghost and the emergence of the quartic one: Working around Minkowski where $N_\mu$ is also taken to be small, the first order bulk equations of motion become
\beq
&\partial_u^2 \t h_{\mu\nu} - 2 \partial_u \partial_{(\mu}N_{\nu)}=0\,,&\\
&\partial_\sigma \partial_u (\t h^\sigma_\mu-\delta^\sigma_\mu \t h)-\partial^\sigma(\partial_\sigma N_\mu - \partial_\mu N_\sigma) = 0\,,&
\eeq
which are solved by $\t h^{(1)}_{\mu\nu}=(1-u) h_{\mu\nu}$ (as in the non-covariant case \eqref{H1}), and a constant $N_\mu$ along the $u$ direction. This linear solution, as before, is sufficient to obtain cubic effective action which consequently contains at most two powers of $N_\mu$, because there are originally two $N_\mu$'s present in the action and $\t h^{(1)}_{\mu\nu}$ does not contain any. This explains the absence of the cubic ghost since there cannot be any cubic self-interaction of the helicity-0 mode which is contained completely in the St\"uckelberg field $N_\mu$ in the decoupling limit \cite{Claudia}. This argument, however, breaks down beyond that order because higher order solutions will necessarily contain $N_\mu$ (e.g. $\t h^{(2)}_{\mu\nu}\supset {N_\mu}^2 $) which upon reduction to 4D result in quartic and higher order terms in $N_\mu$, explaining the presence of the interaction \eqref{pi_4}.

%%%%%%%%%%%%%%%%%%%%%%%%%%%%%%%%%%%%%%%%%%%%%%%%%%%%%%%%%%%%%%%%%%%

\renewcommand{\theequation}{B-\Roman{equation}}
\setcounter{equation}{0} 

\section*{Appendix B. AED with 5D Gauss-Bonnet Term}
In this section we extend the framework of the auxiliary extra dimension (AED) by addition of terms descendent from the Gauss-Bonnet (GB) action. The latter, being the only other five-dimensional Lovelock invariant besides the Einstein-Hilbert term, is given by
\beq
\mathcal{L}_{GB}=\frac{\kappa}{4}\left(R^2-4R_{AB} R^{AB}+R_{ABCD}R^{ABCD}\right),
\label{GB}
\eeq
with $A,\ldots=0,1,2,3,5$ and $\kappa$ being an arbitrary constant.

In order to find the AED analog of \eqref{GB} we impose $g_{\mu 5}=0$ and $g_{55}=1$ on the metric tensor. Furthermore, since we are not interested in generating derivative self-interactions for graviton, we remove terms containing four-dimensional derivatives. As a result \eqref{GB} reduces to
\beq
\mathcal{L}_{GB}^{AED}=\kappa[g^{\mu \nu} \partial_5 k_{\mu \nu}  \left(  k^2-k_{\alpha \beta}^2 \right) +2\partial_5 k_{\mu \nu} \left( k^{\mu \alpha} k_{\alpha}^{\nu}-k^{\mu \nu}k  \right) \nonumber \\
+\frac{1}{4} \left( -14k_{\mu \nu}^4+16 k k_{\mu \nu}^3+7k_{\mu \nu}^2 k_{\alpha \beta}^2-10k^2k_{\mu \nu}^2+k^4 \right) ]\,.
\label{GBAED}
\eeq
However, from the definition of $k_{\mu \nu}$ it follows that there are terms with more than one derivative per field in \eqref{GBAED}, hence, one needs to introduce a boundary term in order for the variation principle to be well-defined. Including those we get the following modification to the Einstein-Hilbert Lagrangian
\beq
\label{L_GB}
\mathcal{L}_{\rm{mass}} = M_{pl}^2m^2\left[ \left.  \frac{\kappa}{3} \sqrt{-g} (2k_{\mu \nu}^3-3k k_{\mu \nu}^2+k^3)\right \vert _{u=0}^{u=1} 
- \int_{0}^{1} du\sqrt{-\tg}(k_{\mu \nu}^2-k^2+\mathcal{L}_{GB}^{AED})\right]\!,\nonumber \\
\eeq
In order to integrate out $u-$dimension, one has to find the solution to the bulk equations of motion which now generalizes to 
\beq
g_{\mu \nu} \partial _u k-\partial_u k_{\mu \nu}=-\frac{1}{2}g_{\mu \nu}(k^2+k_{\alpha \beta}^2)-2k_{\mu \alpha}k^\alpha_\nu+kk_{\mu \nu}\nonumber \\
+\kappa \left[ \partial_u^2 k_{**}k^{**}+\partial_u k_{**}k^{**} k_{**} +k_{**}k^{**} k_{**} k^{**} \right]_{\mu \nu}\,,
\label{eqGB}
\eeq
where we have presented the contribution of the GB term schematically since it is sufficient for our purposes, as will be seen shortly. We choose the boundary condition to be $\tg_{\mu\nu}(x,u=1)=\eta_{\mu\nu}$.

The easiest way of solving \eqref{eqGB} for $\tilde{h}_{\mu \nu}(x,u)\equiv g_{\mu \nu}(x,u)-\eta_{\mu \nu}$ is to proceed order-by-order in four-dimensional metric perturbations $h_{\mu \nu}(x)\equiv \tilde{h}_{\mu \nu}(x,u=0)$. One immediately notices that the newly added terms proportional to $\kappa$ in \eqref{eqGB} start to change the solution only at the $4^{th}$ order in $h_{\mu \nu} (x)$. Using therefore the old linear solution \eqref{H1} it follows that the Gauss-Bonnet term does not contribute to the cubic 4D effective action: The cubic bulk terms in \eqref{GBAED} lead to $\partial_u^2 H^{(1)}=0$, while the boundary terms in \eqref{L_GB} evaluated to 3\rd order are identical at $u=0$ and $u=1$ and cancel each other. 

As in \S\ref{quartic}, to find the 4D action up to 4$^{th}$ order one only needs the 2\nd order bulk solution \eqref{H2} which leads to the following four-dimensional Lagrangian
\beq
\mathcal{L}&=&M^2_{pl}\sqrt{-g}R-\frac{m^2M^2_{pl}}{4}\sqrt{-g}\times (h_{\mu \nu}^2-h^2-h_{\mu \nu}^3+\frac{5}{4} hh_{\mu \nu}^2-\frac{1}{4} h^3\nonumber \\
&&+\frac{1}{144}\left( 6(22+3\kappa)h_{\mu \nu}^4-(53+9\kappa)h_{\mu\nu}^2 h_{\alpha\beta}^2 - 12(11+2\kappa)hh_{\mu\nu}^3\right.\nonumber \\
&&\left. +2(29+9 \kappa)h^2 h_{\mu \nu}^2-(5+3\kappa)h^4 \right)+O(h^5)),
\eeq
with indices contracted by the Minkowsi metric $\eta^{\mu \nu}$. It is easy to see that this will never match the tuned potential of \cite{Giga_FP} for any value of $\kappa$, meaning that the quartic ghost-like pathology of the original model can not be cured by terms of geometrical origin, since the GB term is the only available one in 5D.

So far we limited ourselves to potentials which are motivated by some 5D geometrical construct, however by giving up that requirement one can naturally generalize the potential term in \eqref{lagr} to 
\beq
\label{gen_pot}
V(g_{\mu\nu})= {m^2}\int_{0}^{+1}du \sqrt{\t g}
\left (  k_{\mu\nu}^2 - k^2 +a_1k_{\mu\nu}^3+a_2kk_{\mu\nu}^2+a_3k^3+\dots\right)\,,
\eeq
where now the coefficients at each order should be chosen such that after reduction to 4D and introduction of St\"uckelberg fields pure helicity-0 interactions add up to a total derivative at each order.

One interesting observation that follows from the $N_\mu$ covariantization (appendix A) is that the n$^{\text{th}}$ order terms in $k_{\mu\nu}$ make no contribution to the (n+1)$^{\text{th}}$ order ghost-like interactions. This is because as was the case for the cubic effective action in the original model, only the first order solution is needed to be substituted in $(k_{\mu\nu})^n$ terms. After covariantization (i.e. replacing $k_{\mu\nu}$ with $K_{\mu\nu}$), $H^{(1)}_{\mu\nu}$ remains independent of $N_\mu$ and therefore the highest power of $N_\mu$ at (n+1)\st order is $(N_\mu)^n$. This, however, cannot possibly affect (n+1)\st order self-interaction of helicity-0 mode.

It is also worth mentioning that the equivalence between one and several spherically symmetric auxiliary dimensions does not survive general modifications of the bulk action that include higher powers of $k_{\mu\nu}$. The addition of $k_{\mu\nu}^n$ terms causes the multi-dimensional model to deviate from its co-dimension one counterpart at (n+1)\st order in perturbations. Similarly the Gauss-Bonnet terms lift this degeneracy because of containing higher powers of $\partial_r$, moreover there is one new GB term for each extra dimension which can in principle be included in the action.

%%%%%%%%%%%%%%%%%%%%%%%%%%%%%%%%%%%%%%%%%%%%%%%%%%%%%%%%%%%%%%

\renewcommand{\theequation}{C-\Roman{equation}}
\setcounter{equation}{0} 

\section*{Appendix C. Fine-Tuning of the Boundary}

There are two sets of coefficients \eqref{expansion} that do not give rise to the ghost. One of them is given by
\beq
b^{(1)}_1&=&0,\qquad \forall~ b^{(1)}_2\neq1,\nonumber \\
b^{(2)}_1&=&-b^{(2)}_2=\frac{b^{(2)}_3}{3}+\frac{1}{24} \left(1-4c_3-2b^{(1)}_2+4c_3b^{(1)}_2+{b^{(1)}_2}^2\right),\quad\forall~ b^{(2)}_3,\nonumber\\
b^{(2)}_4&=&\frac{1}{4}(b^{(1)}_2-1)(-1+4c_3+2b^{(1)}_2),\nonumber\\
b^{(3)}_1&=&\frac{b^{(3)}_6}{3}+\frac{1}{48}(8c_3^2(b^{(1)}_2-1)-(b^{(1)}_2-1)(-3+16d_5+3b^{(1)}_2)-4(3+4b^{(1)}_2)b^{(2)}_3\nonumber\\
&&+4c_3(4+(-5+b^{(1)}_2)b^{(1)}_2+4b^{(2)}_3))\nonumber\\
b^{(3)}_2&=&\frac{{b^{(2)}_3}^2}{18(b^{(1)}_2-1)}+\frac{1}{864}((b^{(1)}_2-1)(-96c_3^2+432d_5+12c_3(27-5b^{(1)}_2)\nonumber\\
&&+(b^{(1)}_2-1)(61+5b^{(1)}_2))+24(9-10c_3+16b^{(1)}_2)b^{(2)}_3+288(b^{(3)}_5-b^{(3)}_6))\nonumber\\
b^{(3)}_3&=&-\frac{b^{(3)}_5}{3}-\frac{{b^{(2)}_3}^2}{18(b^{(1)}_2-1)}-\frac{1}{864}((b^{(1)}_2-1)(48c_3^2+144d_5\nonumber\\
&&+12c_3(3+b^{(1)}_2)+(b^{(1)}_2-1)(7+5b^{(1)}_2))+48(c_3+2b^{(1)}_2)b^{(2)}_3)\nonumber
\eeq
\beq
b^{(3)}_4&=&\frac{2{b^{(2)}_3}^2}{3(-1+b^{(1)}_2)}+\frac{1}{72}((b^{(1)}_2-1)(84c_3^2+108d_5+12c_3b^{(1)}_2+(b^{(1)}_2-1)(1+2b^{(1)}_2))\nonumber\\
&&+6b^{(2)}_3(-9+20c_3+4b^{(1)}_2)),\nonumber\\
b^{(3)}_7&=&\frac{b^{(1)}_2-1}{24}(7-72d_5-14b^{(1)}_2+4(-3c_3(3+c_3)+6c_3b^{(1)}_2+{b^{(1)}_2}^2)),\quad \forall~ b^{(3)}_{5,6}.\nonumber
\eeq
while the other one being
\beq
b^{(1)}_1&=&\frac{1}{2}(1-b^{(1)}_2),\qquad \forall~ b^{(1)}_2\neq1,\nonumber \\
b^{(2)}_1&=&-\frac{b^{(2)}_3}{3}-\frac{b^{(1)}_2-1}{24}(-8+16c_3+3b^{(1)}_2),\quad\forall~ b^{(2)}_3,\nonumber\\
b^{(2)}_2&=&\frac{1}{12}((-1+2c_3)(-1+b^{(1)}_2)-2b^{(2)}_3),\nonumber\\
b^{(2)}_4&=&\frac{b^{(1)}_2-1}{4}(-1+4c_3+2b^{(1)}_2),\nonumber\\
b^{(3)}_1&=&\frac{1}{48}(4(b^{(1)}_2-1)(c_3^2+22d_5)-4c_3(13+b^{(1)}_2(-16+3b^{(1)}_2)+4b^{(2)}_3)\nonumber\\
&&+b^{(1)}_2(-23+b^{(1)}_2(5+4b^{(1)}_2)+16b^{(2)}_3)+2(7+6b^{(2)}_3-8b^{(3)}_6)),\nonumber\\
b^{(3)}_2&=&-\frac{7{b^{(2)}_3}^2}{18(-1+b^{(1)}_2)}+\frac{1}{864}((1-b^{(1)}_2)(-85+408c_3^2+1080d_5+12c_3(27+7b^{(1)}_2)\nonumber\\
&&+b^{(1)}_2(14+53b^{(1)}_2))-12(-9+40c_3+20b^{(1)}_2)b^{(2)}_3-144(2b^{(3)}_5+b^{(3)}_6)),\nonumber
\eeq
\beq
b^{(3)}_3&=&\frac{{b^{(2)}_3}^2}{18(-1+b^{(1)}_2)}+\frac{1}{864}((b^{(1)}_2-1)(48c_3^2+144d_5+12c_3(3+b^{(1)}_2)\nonumber\\
&&+(b^{(1)}_2-1)(7+5b^{(1)}_2))+24b^{(2)}_3(2c_3+b^{(1)}_2)-144b^{(3)}_5),\nonumber\\
b^{(3)}_4&=&\frac{2{b^{(2)}_3}^2}{3(-1+b^{(1)}_2)}+\frac{1}{72}((b^{(1)}_2-1)(84c_3^2+108d_5+12c_3b^{(1)}_2+(b^{(1)}_2-1)(1+2b^{(1)}_2))\nonumber\\
&&+6b^{(2)}_3(-9+20c_3+4b^{(1)}_2)),\nonumber\\
b^{(3)}_7&=&\frac{b^{(1)}_2-1}{24}(7-72d_5-14b^{(1)}_2+4(-3c_3(3+c_3)+6c_3b^{(1)}_2+{b^{(1)}_2}^2)),\quad \forall~ b^{(3)}_{5,6}.\nonumber
\eeq
These coefficients give the most general boundary condition at $u=1$, for which the theory is ghost-free (in decoupling limit) up-to 4$^{\text{th}}$ order. It is quite straightforward to continue this tuning to arbitrary order.

\end {document}